\documentclass[11pt,twoside]{article}
\usepackage{./asp2010}
\usepackage{graphicx}

\resetcounters

\begin{document}

\title{The Age and Helium Abundance of the Galactic Bulge}
\author{David M. Nataf$^1$
\affil{$^1$Research School of Astronomy and Astrophysics, Australian National University, Canberra, ACT 2611, Australia}}

\begin{abstract}
I discuss the age and helium abundance of the Galactic bulge stellar population.  I present examples as to how age and helium abundance can be degenerate observationally, and thus, how unstated assumptions of the helium-metallicity relation can result in biased age-metallicity determinations. I summarise efforts to tackle this degeneracy using the red giant branch bump, forward modelling of microlensing selection effects, and direct mass and metallicity measurements of stars along the red giant branch. I make the case that the degeneracy between these two parameters can be resolved in the near future.
\end{abstract}

\section{Introduction}
The Galactic bulge comprises a substantial fraction of the Milky Way's stellar mass, as well as the bulk of both the Galaxy's oldest and most metal-rich stars. The age distribution of the Galactic bulge is thus a fundamental parameter of Galactic formation and evolution and needs to be understood before a consistent archeological history of the Milky Way can emerge. HST colour-magnitude diagrams of the Galactic bulge with disc contamination removed by means of proper motion criteria show a main-sequence turnoff that is as dim relative to the horizontal branch as seen toward globular clusters \citep{2002AJ....124.2054K}, indicating a very old age ($t\geq 10$ Gyr). This argument, and others such as the chemical enrichment patterns of bulge stars, led to a quasi-consensus in the literature of a purely old Milky Way bulge:
\begin{itemize}
\item ``the CMD of BaadeÕs Window field indicate a uniformly old age for stars in the Galactic bulge, thus helping to settle the question of the formation of galactic bulges." - \citet{1995ASPC...86..305R}
\item ``The bulge of our Galaxy formed at the same time and even faster than the inner Galactic halo." - \citet{1999Ap&SS.265..311M}
\item ``The population with nondisk kinematics (which we conclude to be the bulge) has an old main-sequence turnoff point, similar to those found in old, metal-rich bulge globular clusters." - \citet{2002AJ....124.2054K}
\item  ``the bulge age, which we found to be as large as that of Galactic globular clusters, or $\gtrsim$ 10 Gyr. No trace is found for any younger stellar population." - \citet{2003A&A...399..931Z}
\end{itemize}

This consensus has recently been undermined by spectroscopic observations of microlensed stars towards the bulge, which show temperatures for main-sequence-turnoff stars that are too high relative to expectations for a purely old stellar population, as well as surface gravities on the subgiant branch that are too low. \cite{2013A&A...549A.147B}. Evidence for both a purely old stellar population and a mixed population with stars as young as $\sim$3 Gyr abounds from other sources as well, such as the long periods of Galactic bulge Mira variables \citep{1991MNRAS.248..276W}. The assumption of a purely old bulge with canonical enrichment trends was also found to be inconsistent with the luminosity function of the red giant branch \citep{2011ApJ...730..118N}.

Though the bulge may be worthy of precision characterisation, such a description remains elusive, and the age distribution is one of the most important mysteries. In this conference proceeding I make the case that the degeneracy between age and helium abundance is likely contributing to our misunderstanding, that the helium abundance may be non-standard for the more metal-rich stars, and I discuss three observational probes that I argue render this difficult problem tractable.  

\section{The Red Giant Branch Bump}
The red giant branch bump (RGBB) is an excess in the luminosity function  of first-ascent red giant stars that approximately occurs when the hydrogen-burning shell meets the convective zone \citep{1997MNRAS.285..593C}. The magnitude and normalisation of the RGBB are steeply sensitive to a stellar population's metallicity, age, and helium abundance \citep{1997MNRAS.285..593C,2010ApJ...712..527D,2011ApJ...730..118N}. 

\begin{figure}[!ht]
\begin{center}
\includegraphics[scale=0.33]{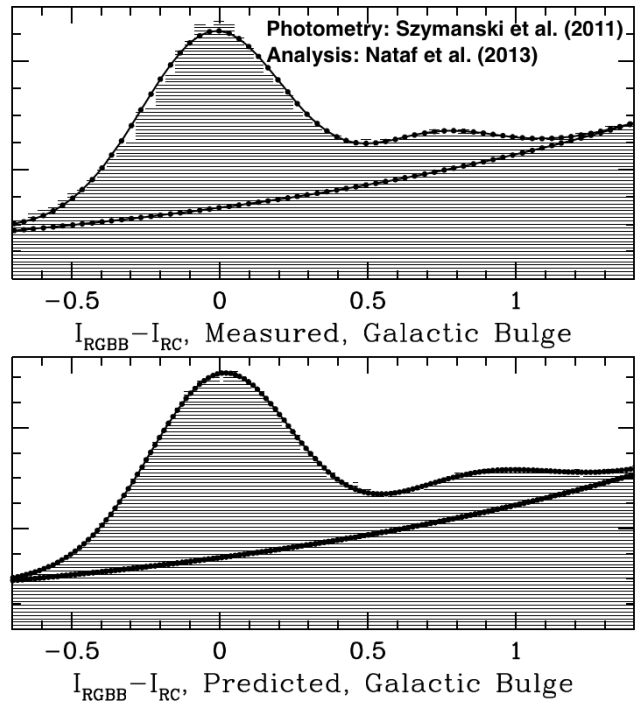}
\caption{Comparison of the observed red giant luminosity function for the Galactic bulge (Top) relative to one simulated with an empirical calibration from 72 Galactic globular clusters (Bottom). The observed RGBB has a smaller normalisation and doesn't extend as far to fainter magnitudes.  }
\end{center}
\label{Fig:NatafFig1}
\end{figure}

\begin{figure}[!ht]
\begin{center}
\includegraphics[scale=0.33]{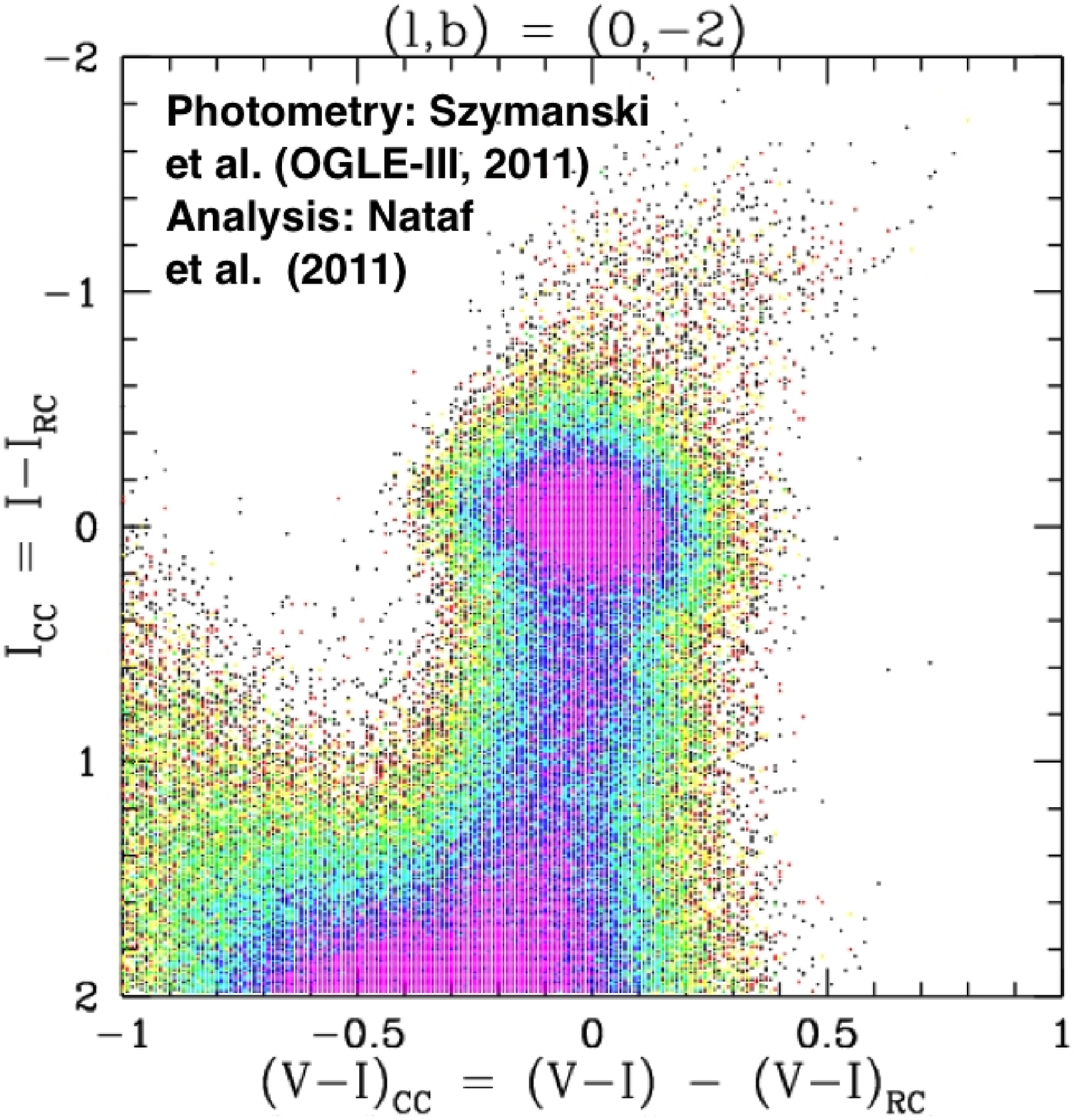}
\caption{The CCCMD toward a bulge sightline from OGLE-III photometry. Regions of the CMD with the greatest density of stars are shown in magenta. }
\end{center}
\label{Fig:NatafFig2}
\end{figure}

However, though the Galactic bulge RGBB could in principle be a potent probe, it had not been measured until recently. The high differential reddening toward the bulge made it difficult to adequately ascertain this feature of the bulge color-magnitude diagram. \citet{2011ApJ...730..118N} used deep OGLE-III photometry \citep{2011AcA....61...83S} to construct clump-centric color-magnitude (CCCMD) diagrams of the bulge, in a manner that would mitigate the impact of differential reddening. A CCCMD is shown in Figure \ref{Fig:NatafFig1}. This allowed a first measurement of the Galactic bulge RGBB. \citet{2013ApJ...766...77N} followed-up with an empirical calibration with metallicity-dependence of RGBB brightness and number counts spanning 72 Galactic globular clusters observed by the \textit{Hubble Space Telescope (HST)}. This calibration was applied to a measurement of the Galactic bulge RGBB with a more precise accounting of differential reddening toward a sightline ($(l,b) = (0,-2)$) with relatively low geometric dispersion. It was shown that the magnitude distribution of the RGBB was $\sim$0.10 mag too bright in the mean, with a normalisation $\sim$20\% too small, relative to expectations from the globular cluster calibration. The predicted and observed luminosity functions are shown in Figure \ref{Fig:NatafFig1}.  Updated YREC stellar models \citep{2012ApJ...746...16V} were used to show that both a $\sim$40\% reduction in age or an increase in the initial helium abundance of ${\Delta}Y=+0.06$, could explain the change in the number counts. However, the age variation would brighten the RGBB by $\sim$0.30 mag, whereas the helium variation would only brighten it by $\sim$0.10 magnitudes. 

Thus, non-standard helium with an age distribution similar to the globular clusters is a better match to the observations than either the assumption of standard helium enrichment with a much younger age for the bulge than that of globular clusters, or the assumption of both an old age and standard helium enrichment for the bulge. 

\section{The Selection Function of Microlensed Stars Toward the Galactic Bulge}

The microlensed stars observed by \citet{2013A&A...549A.147B} have measured spectroscopic parameters that are inconsistent with the assumption of a purely old bulge that can be studied with standard stellar models. The main-sequence turnoff temperatures are often too high, and the subgiant surface gravities are often too low, indicative of a stellar population with $\sim$20\% of its stars younger than $\sim$5 Gyr and $\sim$30\% younger than 7 Gyr. 

This result is at odds with the prior observational and theoretical consensus, and it is thus important to investigate which factors other than age, if any, could contribute to this discrepancy. \citet{2012ApJ...751L..39N} suggested enhanced helium enrichment as a possible cause. They used Dartmouth stellar models \citep{2008ApJS..178...89D} to show that the application of standard isochrones to determine the spectroscopic ages of a helium-enhanced population would lead to systematically underestimated ages. The bias is expected to be highest on the subgiant branch as well as for lower-mass main-sequence stars that have not yet evolve to the turnoff, and lowest near the main-sequence turnoff. This is observed for the subset of \citet{2013A&A...549A.147B} with [Fe/H] $\geq 0.0$, but the number counts are simply too small to state that the variation in the age offset is statistically significant. \citet{2012ApJ...751L..39N} found a $\sim 2 \sigma$ preference for ${\Delta}Y=+0.098$. 

Given the potential diagnostic power of the microlensing sample, it is important to understand the selection function, such as any possible bias toward younger stars or the rate of disc contamination. Collaborators and I have been working to construct various bulge synthetic populations to investigate how the observed distribution of microlensed stars could differ from the intrinsic distributions. We first used the Besancon Galaxy model \citep{2003A&A...409..523R} and found that the predicted rate of disc contamination is $\sim$8\%. These disc stars are not expected to contribute young, metal-rich stars, but rather old stars with [Fe/H] $\approx -0.50$, as they are in fact thick disc stars at the other end of the Galaxy, with a typical distance of $\sim$600 pc separating them from the Galactic plane.  There is a (small) bias due to disc contamination, but it has the exact opposite effect necessary to reconcile the discrepancy between photometric and spectroscopic turnoff ages. 

We also tested for an synthetic population constructed with an arbitrarily elevated fraction of young stars using the Dartmouth stellar database \citep{2008ApJS..178...89D}. We found that there was indeed a bias toward younger stars, but it is not nearly large enough to reconcile the microlensing sample with canonical expectations. The fraction of stars younger than t$=5$ Gyr does increase, by a factor of $\sim$40\%. This would imply that the 23\% young star fraction measured by \citet{2013A&A...549A.147B} is really indicative of a $\sim$16\% young star fraction, which is still an order-of-magnitude higher than canonical predictions of no more than a trace population younger than $\sim$10 Gyr. 

\begin{figure}[!ht]
\plotone{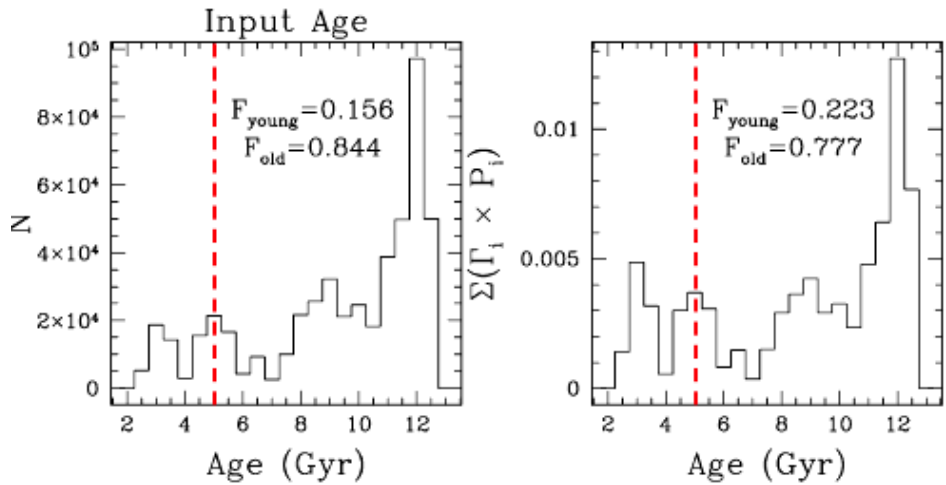}
\caption{The effect of selection effects on the age distribution of microlensed dwarf stars toward the Galactic bulge. In this simulation, the fraction of stars younger than 5 Gyr is increased from 15.6\% to 22.3\%.}
\label{Fig:NatafFig3}
\end{figure}

We are currently moving forward with more rigorous testing, where we use the \citet{2003ApJ...592..172H} Galactic dynamics model, the chemical evolution model of \citet{2012A&A...548A..60G}, and the Dartmouth stellar isochrone database of \citet{2008ApJS..178...89D} to account for the bulk of the selection effects. The blue straggler population, which appears younger than it really is, will be normalised using the investigation of \citet{2011ApJ...735...37C}. If selection effects and blue straggler contamination are ruled out as explanations, the two remaining explanations will be a failure of stellar evolution models as diagnostics of the bulge stellar population, or recognition that the bulge indeed has a substantial young population.

\section{Detached Red Giant Eclipsing Binaries}
Age and helium abundance are degenerate virtually everywhere on the colour-magnitude diagram. Further, both are often degenerate with other factors such as the quantity of reddening, the reddening curve, the distance, the metallicity distribution function, and the binary fraction, which makes them extremely difficult to disentangle. Given this difficulty, any additional independent observational probe has great potential to be of use, and one such probe is the mass-metallicity relation of stars in the red giant phase. 

\citet{2012AcA....62...33N} analysed a sample of $\sim$10,000 eclipsing binaries with parameters determined by \citet{2005ApJ...628..411D}. They found $\sim$300 systems that might be detached red giant eclipsing binary twins. For these systems, with median periods of $\sim$6 days and thus typical radial velocity amplitudes of $\sim$100 km/s, the masses and metallicities could be measured in a relatively straightforward way. 

To gauge how the masses and metallicities could be interpreted, \citet{2012AcA....62...33N} used an updated version of the Yale Rotating Evolution Code \citep{2012ApJ...746...16V} to make the follow prediction for masses of stars near the base of the red giant branch:
\begin{equation}
\log\biggl( \frac{M}{M_{\odot}}   \biggl)\,= 0.026 + 0.126\rm{[M/H]} - 0.276\log \biggl(  \frac{t}{10\,Gyr}  \biggl) - 0.937(Y-0.27),
\end{equation}
where [M/H] is the metallicity, $t$ is the age, and $Y$ is the initial helium abundance. As with other methods, there are degenracies, but these are not the same degeneracies. The metallicity can in principle be factored out. Further, as a lower mass at fixed metallicity implies either a higher helium abundance or an older age. This degeneracy has the opposite sign to that of spectroscopic investigations of the main-sequence turnoff and subgiant branch, where higher helium abundance resembles a younger age. 

The sensitivity is also quite large: An increase in the initial helium abundance of a single percentage point at fixed age and metallicity leads to a 2\% reduction in the mass. In contrast, masses for red giant eclipsing binary twins can be determined to better than 1\% \citep{2012ApJ...750..144G}, and the actual helium offset relative to standard isochrones might be substantially larger than a single percentage point, particularly at the highest metallicities. The prospect for a very tight constraint is there. Further, the sample of candidates should increase by a factor of $\sim$10 once OGLE-III time series photometry becomes available, as the spatial coverage is $\sim$10 higher than in OGLE-II, with greater sampling and more precise photometry as well. 

A proposal was submitted to follow-up some of these targets on the Omega multi-object spectrograph, which resulted in $\sim$50 hours of observations. We focused on a single field with $\sim$ 80 candidates centred on $(\alpha, \delta) = (18:02:39.20, -28:55:10.26)$, corresponding to $(l,b)=(1.90, -3.21)$. 

\begin{figure}[!ht]
\plotone{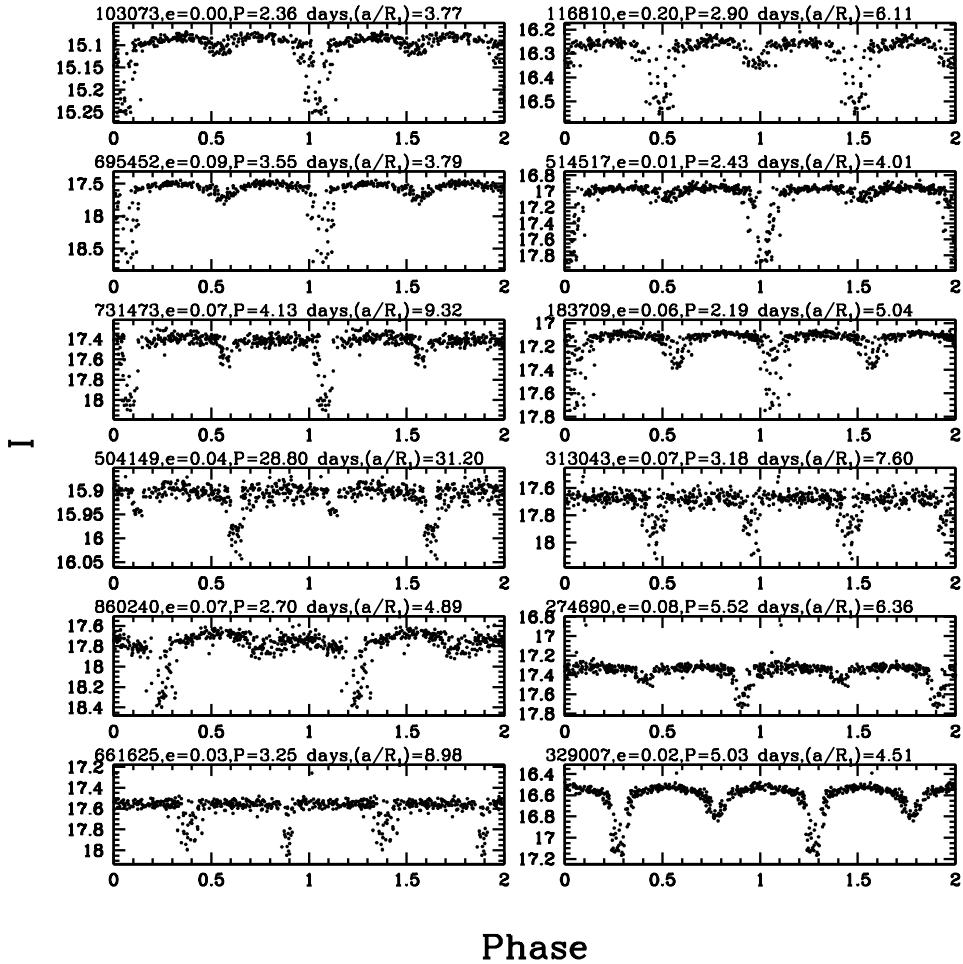}
\caption{A sample of eclipsing binary lightcurves that have been followed-up with AAT time-series spectroscopy. }
\label{Fig:NatafFig4}
\end{figure}

\begin{figure}[!ht]
\plotone{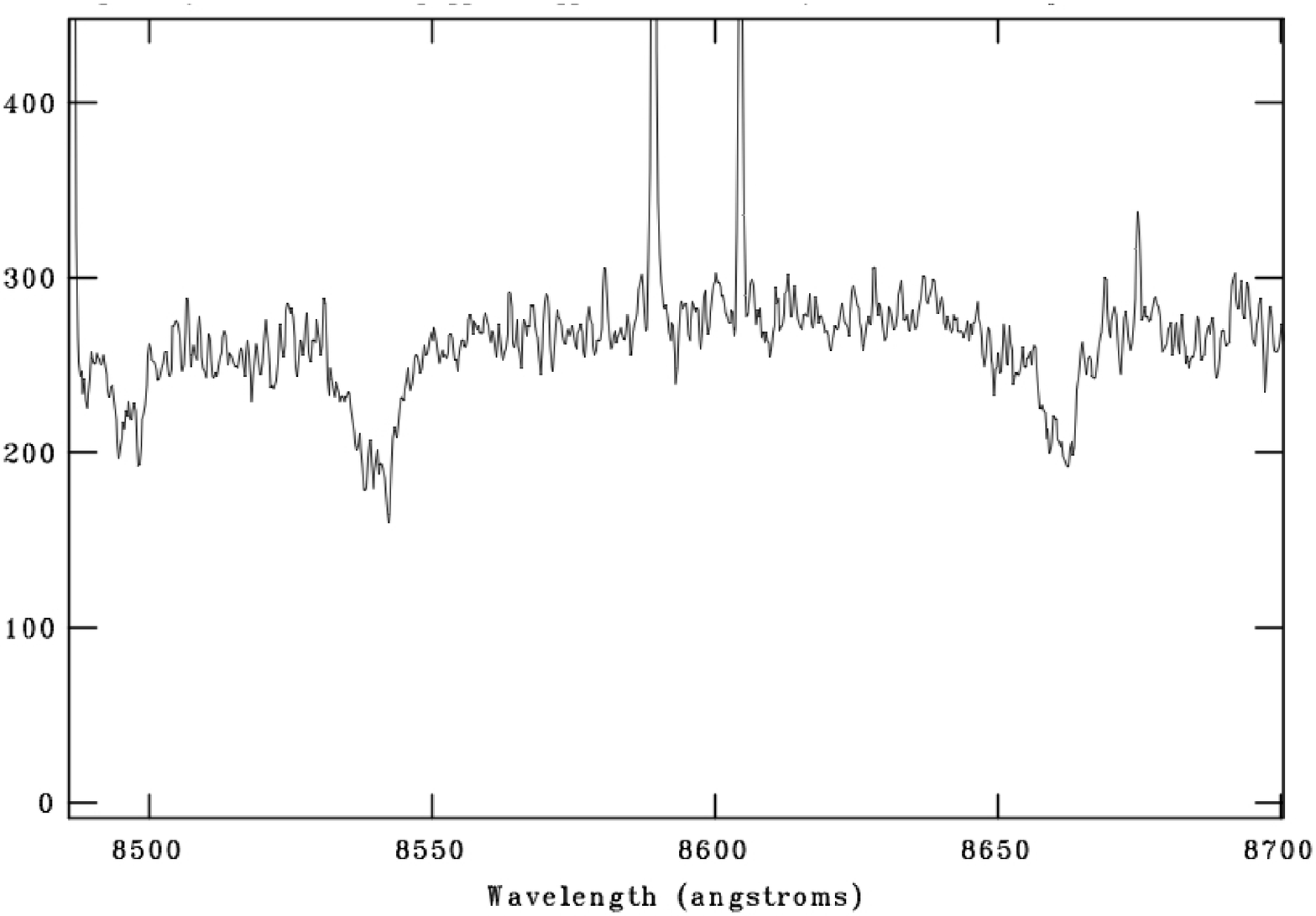}
\caption{The partially-reduced spectra of a detached red giant star in an eclipsing binary with a spectrum indicating a likely double-lined calcium triplet.}
\label{Fig:NatafFig5}
\end{figure}

Data reduction and analysis is now in a preliminary stage. Careful consideration will need to be given to issues of how to deal with the binary partners, and how many of them are red giant twins, subgiant stars, or turnoff stars. I show a sample of 12 eclipsing binaries with mean densities and colours consistent with a red giant morphology in Figure \ref{Fig:NatafFig4}. In Figure \ref{Fig:NatafFig5}, I show the partially-reduced spectra of a candidate red giant in an eclipsing binary, which appears to show a double-lined calcium triplet. 

\section{Conclusions}
A precise accounting of both the mean age of the Galactic bulge and its mapping onto the metallicity distribution function has thus far remained elusive, as different methodologies have yielded different conclusions. The brightness and number counts of the Galactic bulge RGBB \citep{2011ApJ...730..118N,2013ApJ...766...77N} and the microlensing sample of \citet{2013A&A...549A.147B} are both inconsistent with the prior consensus that the bulge is a purely old population (stated assumption) that can be analysed with stellar models that assume standard helium-enrichment (unstated assumption). At this time, the brightness and number counts of the RGBB suggest ${\Delta}Y=+0.06$ in the mean \citep{2013ApJ...766...77N}, and the turnoff age discrepancy can be completely eliminated by having ${\Delta}Y\approx+0.11$ at [Fe/H]$\approx$0.30. \citep{2012ApJ...751L..39N}. However, both these estimates are limiting values, computed under the assumption of no age spread. 

The degeneracy between the age and helium-enrichment of any stellar population, let alone one such as the bulge with an uncertain distance distribution, reddening distribution, binary distribution, and metallicity distribution, is extremely difficult to disentangle. It is my hope that a more precise quantisation of the selection effects in the microlensing sample, and a mapping of the mass-metallicity relation of Galactic bulge red giant stars, will help elucidate this issue.

\acknowledgements I would like to thank Andrew Gould, Martin Asplund, Marc H. Pinsonneault, Mathew Penny, Calen Henderson, Andrzej Udalski, Jennifer Johnson, and Krzysztof Stanek for their contributions at various points in the development of this work. I also thank the OGLE collaboration, whose spectacular data made the majority of this research possible. 


\begin{thebibliography}{}
\expandafter\ifx\csname natexlab\endcsname\relax\def\natexlab#1{#1}\fi
\expandafter\ifx\csname url\endcsname\relax
  \def\url#1{\texttt{#1}}\fi
\expandafter\ifx\csname urlprefix\endcsname\relax\def\urlprefix{URL }\fi
\providecommand{\eprint}[2][]{\url{#2}}

\bibitem[{{Bensby} et~al.(2013){Bensby}, {Yee}, {Feltzing}, {Johnson}, {Gould},
  {Cohen}, {Asplund}, {Mel{\'e}ndez}, {Lucatello}, {Han}, {Thompson},
  {Gal-Yam}, {Udalski}, {Bennett}, {Bond}, {Kohei}, {Sumi}, {Suzuki}, {Suzuki},
  {Takino}, {Tristram}, {Yamai}, \& {Yonehara}}]{2013A&A...549A.147B}
{Bensby}, T., {Yee}, J.~C., {Feltzing}, S., {Johnson}, J.~A., {Gould}, A.,
  {Cohen}, J.~G., {Asplund}, M., {Mel{\'e}ndez}, J., {Lucatello}, S., {Han},
  C., {Thompson}, I., {Gal-Yam}, A., {Udalski}, A., {Bennett}, D.~P., {Bond},
  I.~A., {Kohei}, W., {Sumi}, T., {Suzuki}, D., {Suzuki}, K., {Takino}, S.,
  {Tristram}, P., {Yamai}, N., \& {Yonehara}, A. 2013, \aap, 549, A147.
  \eprint{1211.6848}

\bibitem[{{Cassisi} \& {Salaris}(1997)}]{1997MNRAS.285..593C}
{Cassisi}, S., \& {Salaris}, M. 1997, \mnras, 285, 593.
  \eprint{arXiv:astro-ph/9702029}

\bibitem[{{Clarkson} et~al.(2011){Clarkson}, {Sahu}, {Anderson}, {Rich},
  {Smith}, {Brown}, {Bond}, {Livio}, {Minniti}, {Renzini}, \&
  {Zoccali}}]{2011ApJ...735...37C}
{Clarkson}, W.~I., {Sahu}, K.~C., {Anderson}, J., {Rich}, R.~M., {Smith},
  T.~E., {Brown}, T.~M., {Bond}, H.~E., {Livio}, M., {Minniti}, D., {Renzini},
  A., \& {Zoccali}, M. 2011, \apj, 735, 37. \eprint{1105.4176}

\bibitem[{{Devor}(2005)}]{2005ApJ...628..411D}
{Devor}, J. 2005, \apj, 628, 411. \eprint{arXiv:astro-ph/0504399}

\bibitem[{{Di Cecco} et~al.(2010){Di Cecco}, {Bono}, {Stetson}, {Pietrinferni},
  {Becucci}, {Cassisi}, {Degl'Innocenti}, {Iannicola}, {Prada Moroni},
  {Buonanno}, {Calamida}, {Caputo}, {Castellani}, {Corsi}, {Ferraro},
  {Dall'Ora}, {Monelli}, {Nonino}, {Piersimoni}, {Pulone}, {Romaniello},
  {Salaris}, {Walker}, \& {Zoccali}}]{2010ApJ...712..527D}
{Di Cecco}, A., {Bono}, G., {Stetson}, P.~B., {Pietrinferni}, A., {Becucci},
  R., {Cassisi}, S., {Degl'Innocenti}, S., {Iannicola}, G., {Prada Moroni},
  P.~G., {Buonanno}, R., {Calamida}, A., {Caputo}, F., {Castellani}, M.,
  {Corsi}, C.~E., {Ferraro}, I., {Dall'Ora}, M., {Monelli}, M., {Nonino}, M.,
  {Piersimoni}, A.~M., {Pulone}, L., {Romaniello}, M., {Salaris}, M., {Walker},
  A.~R., \& {Zoccali}, M. 2010, \apj, 712, 527. \eprint{1002.2074}

\bibitem[{{Dotter} et~al.(2008){Dotter}, {Chaboyer}, {Jevremovi{\'c}},
  {Kostov}, {Baron}, \& {Ferguson}}]{2008ApJS..178...89D}
{Dotter}, A., {Chaboyer}, B., {Jevremovi{\'c}}, D., {Kostov}, V., {Baron}, E.,
  \& {Ferguson}, J.~W. 2008, \apjs, 178, 89. \eprint{0804.4473}

\bibitem[{{Graczyk} et~al.(2012){Graczyk}, {Pietrzy{\'n}ski}, {Thompson},
  {Gieren}, {Pilecki}, {Udalski}, {Soszy{\'n}ski}, {Ko{\l}aczkowski},
  {Kudritzki}, {Bresolin}, {Konorski}, {Mennickent}, {Minniti}, {Storm},
  {Nardetto}, \& {Karczmarek}}]{2012ApJ...750..144G}
{Graczyk}, D., {Pietrzy{\'n}ski}, G., {Thompson}, I.~B., {Gieren}, W.,
  {Pilecki}, B., {Udalski}, A., {Soszy{\'n}ski}, I., {Ko{\l}aczkowski}, Z.,
  {Kudritzki}, R.-P., {Bresolin}, F., {Konorski}, P., {Mennickent}, R.,
  {Minniti}, D., {Storm}, J., {Nardetto}, N., \& {Karczmarek}, P. 2012, \apj,
  750, 144. \eprint{1203.2517}

\bibitem[{{Grieco} et~al.(2012){Grieco}, {Matteucci}, {Pipino}, \&
  {Cescutti}}]{2012A&A...548A..60G}
{Grieco}, V., {Matteucci}, F., {Pipino}, A., \& {Cescutti}, G. 2012, \aap, 548,
  A60. \eprint{1209.4462}

\bibitem[{{Han} \& {Gould}(2003)}]{2003ApJ...592..172H}
{Han}, C., \& {Gould}, A. 2003, \apj, 592, 172. \eprint{arXiv:astro-ph/0303309}

\bibitem[{{Kuijken} \& {Rich}(2002)}]{2002AJ....124.2054K}
{Kuijken}, K., \& {Rich}, R.~M. 2002, \aj, 124, 2054.
  \eprint{arXiv:astro-ph/0207116}

\bibitem[{{Matteucci} \& {Romano}(1999)}]{1999Ap&SS.265..311M}
{Matteucci}, F., \& {Romano}, D. 1999, \apss, 265, 311

\bibitem[{{Nataf} et~al.(2012){Nataf}, {Gould}, \&
  {Pinsonneault}}]{2012AcA....62...33N}
{Nataf}, D.~M., {Gould}, A., \& {Pinsonneault}, M.~H. 2012, Acta Astronomica,
  62, 33. \eprint{1203.5791}

\bibitem[{{Nataf} \& {Gould}(2012)}]{2012ApJ...751L..39N}
{Nataf}, D.~M., \& {Gould}, A.~P. 2012, \apjl, 751, L39. \eprint{1112.1072}

\bibitem[{{Nataf} et~al.(2013){Nataf}, {Gould}, {Pinsonneault}, \&
  {Udalski}}]{2013ApJ...766...77N}
{Nataf}, D.~M., {Gould}, A.~P., {Pinsonneault}, M.~H., \& {Udalski}, A. 2013,
  \apj, 766, 77. \eprint{1109.2118}

\bibitem[{{Nataf} et~al.(2011){Nataf}, {Udalski}, {Gould}, \&
  {Pinsonneault}}]{2011ApJ...730..118N}
{Nataf}, D.~M., {Udalski}, A., {Gould}, A., \& {Pinsonneault}, M.~H. 2011,
  \apj, 730, 118. \eprint{1011.4293}

\bibitem[{{Renzini}(1995)}]{1995ASPC...86..305R}
{Renzini}, A. 1995, in Fresh Views of Elliptical Galaxies, edited by
  A.~{Buzzoni}, A.~{Renzini}, \& A.~{Serrano}, vol.~86 of Astronomical Society
  of the Pacific Conference Series, 305

\bibitem[{{Robin} et~al.(2003){Robin}, {Reyl{\'e}}, {Derri{\`e}re}, \&
  {Picaud}}]{2003A&A...409..523R}
{Robin}, A.~C., {Reyl{\'e}}, C., {Derri{\`e}re}, S., \& {Picaud}, S. 2003,
  \aap, 409, 523

\bibitem[{{Szyma{\'n}ski} et~al.(2011){Szyma{\'n}ski}, {Udalski},
  {Soszy{\'n}ski}, {Kubiak}, {Pietrzy{\'n}ski}, {Poleski}, {Wyrzykowski}, \&
  {Ulaczyk}}]{2011AcA....61...83S}
{Szyma{\'n}ski}, M.~K., {Udalski}, A., {Soszy{\'n}ski}, I., {Kubiak}, M.,
  {Pietrzy{\'n}ski}, G., {Poleski}, R., {Wyrzykowski}, {\L}., \& {Ulaczyk}, K.
  2011, Acta Astronomica, 61, 83. \eprint{1107.4008}

\bibitem[{{van Saders} \& {Pinsonneault}(2012)}]{2012ApJ...746...16V}
{van Saders}, J.~L., \& {Pinsonneault}, M.~H. 2012, \apj, 746, 16.
  \eprint{1108.2273}

\bibitem[{{Whitelock} et~al.(1991){Whitelock}, {Feast}, \&
  {Catchpole}}]{1991MNRAS.248..276W}
{Whitelock}, P., {Feast}, M., \& {Catchpole}, R. 1991, \mnras, 248, 276

\bibitem[{{Zoccali} et~al.(2003){Zoccali}, {Renzini}, {Ortolani}, {Greggio},
  {Saviane}, {Cassisi}, {Rejkuba}, {Barbuy}, {Rich}, \&
  {Bica}}]{2003A&A...399..931Z}
{Zoccali}, M., {Renzini}, A., {Ortolani}, S., {Greggio}, L., {Saviane}, I.,
  {Cassisi}, S., {Rejkuba}, M., {Barbuy}, B., {Rich}, R.~M., \& {Bica}, E.
  2003, \aap, 399, 931. \eprint{arXiv:astro-ph/0210660}

\end{thebibliography}

\end{document}